\documentclass{aa}
\usepackage[varg]{txfonts}
\usepackage{natbib}
\def\aap{A\& A}

\def\apj{ApJ}
\def\apjl{ApJL}
\def\apjs{ApJS}
\def\apss{Ap\&SS}

\def\mnras{MNRAS}

\RequirePackage{color}

\begin{document}

\title{Mass functions from the excursion set model}

\subtitle{Mass functions from the excursion set model}

\author{Nicos Hiotelis\inst{1}
  \and Antonino Del Popolo\inst{2,3,4}
%  \thanks{\emph{Present address:}
%    Department of Computer Science, Purdue University,
%    West Lafayette, IN 47907, USA}
%     \and Robert J. Plemmons\inst{3}
     }

%\offprints{R. Plemmons, \email{plemmons@...}}

\institute{1st  Lyceum of Athens, Ipitou 15, Plaka, 10557,
Athens,  Greece
\and
Institute of Modern Physics, Chinese Academy of Sciences,\\
Post Office Box 31, Lanzhou 730000, Peoples Republic of China
  \and Dipartimento di Fisica e Astronomia, University Of Catania, \\
Viale Andrea Doria 6, 95125, Catania, Italy
  \and INFN sezione di Catania,\\
Via S. Sofia 64, I-95123 Catania, Italy}

\date{Received  / Accepted  }

\abstract

%\date{Received 2 November 1992 / Accepted 7 January 1993}

\abstract {} {We aim to study the stochastic evolution of the smoothed overdensity $\delta$ at scale $S$  of the form $\delta(S)=\int_{0}^S K(S,u)\mathrm{d}W(u)$,  where $K$ is a kernel and $\mathrm{d}W$ is the usual Wiener process. } {For  a Gaussian density field, smoothed by the  top-hat filter, in real space, we used a simple  kernel that gives the correct correlation between scales. A Monte Carlo  procedure was used to construct random walks  and to calculate first crossing distributions and consequently mass functions for a constant barrier.} {We show that the  evolution considered here improves the agreement with the results of N-body simulations relative to analytical approximations which have been proposed from the same problem by other authors. In fact, we show that an evolution which is fully consistent with the ideas of the excursion set model, describes accurately the mass function of dark matter haloes for values of $\nu \leq 1$ and underestimates the number of larger haloes. Finally, we show that a constant threshold of collapse, lower than it is usually used, it is able to produce a mass function which approximates the results of N-body simulations for a variety of redshifts and for a wide range of masses.} {A mass function in good agreement with N-body simulations can be obtained analytically using a lower than usual constant collapse threshold.}

%}
\maketitle

\section{Introduction}

Over the course of the past several decades, cosmologists using a large number of observations came up with a model describing the structure and evolution of the universe, dubbed $\Lambda$CDM model. In this model the Universe is constituted by cold dark matter (CDM), and vacuum energy (represented by the cosmological constant $\Lambda$). This model fits a large number of data \citep{DelPopolo2007,Komatsu2011,DelPopolo2013,DelPopolo2014,Planck2016_XIII}, but suffers from drawbacks on small scales 
{\citep[see][]{DelPopolo2017a}}, 
%for a review)}, 
the fine tuning problem \citep{Weinberg1989,Astashenok2012} and the cosmic coincidence problem.
Another fundamental test that the $\Lambda$CDM model has to pass is to accurately predict the dark matter (DM) haloes distribution (i.e., the halo mass function (MF) \citep[see][]{DelPopolo2007a,Hiotelis2006,Hiotelis2013}). The high mass end of the MF at small redshift ( $z\leq 2$) is very sensitive to cosmological parameters like the Universe matter and dark energy (DE) content ($\Omega_{\rm m}$ and $\Omega_{\Lambda}$), the equation of state of the Universe, $w$, and its evolution \citep{Malekjani2015,Pace2014}. At redshifts higher than {the} previously quoted ones, the MF is of fundamental importance in the study of the reionization history of the universe \citep[e.g.,][]{Furlanetto2006}, quasar abundance \citep[e.g.,][]{Haiman2001a}, and to study the distribution of DM.

\cite{Press1974} (PS) proposed a very simple model based on the assumption of Gaussian distribution of the initial density perturbation, and the spherical collapse model. The quoted approach has the drawback of overpredicting the number of objects at small masses, and underpredicting those at high mass  \citep[e.g.,][]{Jenkins2001,White2002}. The extended-PS formalism, or excursion set approach, \citep{Bond1991,Bower1991,Lacey1993,Gardner2001}, introduced to overcome the quoted problems, was unable to solve them.

Extension of the quoted formalism  \citep{DelPopolo1998,DelPopolo1999,DelPopolo2000,Sheth2001}, moving from the spherical collapse to non-spherical collapse gave much better agreement with  N-body simulations \citep{Sheth1999} (ST). However, a deeper analysis of ST,
%\cite{Sheth1999,Sheth2001}
and \cite{Sheth2001} showed that the ST MF overpredicts the halo number at large masses \citep{Warren2006,Lukic2007, Reed2007, Crocce2010,Bhattacharya2011,Angulo2012, Watson2013}, and when the redshift evolution is studied the situation worsen \citep{Reed2007,Lukic2007, Courtin2011}.

Another important issue is that of the universality of the MF, namely its independence on cosmology and redshift. Several studies \citep[e.g.,][]{Tinker2008,Crocce2010,Bhattacharya2011,Courtin2011,Watson2013} showed that the MF is not universal nor in its $z$ dependence or for different cosmologies.

In the present paper, we want to show how the excursion set approach can be improved to the extent that it can produce a MF in good agreement with N-body simulations like that of \citep{Tinker2008} who showed clear evidences of the MF deviations from universality, calibrated the MF at $z=0$ in the $10^{11}<M<10^{15}~h^{-1}~M_{\odot}$ mass range within 5\%, and found the redshift evolution of the same.

The paper is organized as follows. In Sect. 2, we discuss the stochastic process, and Sect. 3 is devoted to results and discussion.

\section{The stochastic process.}

As already reported, the excursion set model is  based on the ideas of  \cite{Press1974}
%\citep{prsc74}
and on their extensions which are presented in the pioneered works of
%  \citep{pea},
%\citep{boet91}
\cite{Bond1991} and
%\citep{laco93}.
\cite{Lacey1993}.
We will improve on these ideas in this paper but before that we will write some useful relations about the density fields and the smoothing filters which will be used in what follows.\\
 The smoothed density perturbation  at the center of a spherical region is
 \begin{equation}
 \delta(R)=\int W_f(r;R)\hat{\delta} (r)4\pi r^2\mathrm{d}r
 ,\end{equation}
   where $\hat{\delta} (r)$ is the density at distance $r$ from the center of the spherical region and $W_f$ is a smoothing filter. Reducing  $R${,} the variable $\delta$ executes a random walk that depends on the form  of the density field and on  the smoothing filter $W_f$. If the density field is Gaussian, then $\delta$ is a central Gaussian variable and its probability density is given by
 \begin{equation}
  \label{a0}
  p(\delta(R)=x)\mathrm{d}x=\frac{1}{\sqrt{2\pi \sigma^2(R)}}\mathrm{exp}\left[-\frac{x^2}{2\sigma^2(R)}\right]\mathrm{d}x
  .\end{equation}
   For a spherically symmetric filter, the variance at radius $R$ is given by
  \begin{equation}
  \label{a1}
  S(R)\equiv \sigma^2(R)=\frac{1}{2\pi^2}\int_{0}^{\infty}k^2P(k){\widehat{W}_f}^2(k;R)\mathrm{d}k
  ,\end{equation}
  where $\widehat{W}_f$ is the Fourier transform of the filter and $P$ is the power spectrum.\\
  The correlation of values of $\delta$ between scales is given by the autocorrelation function that is
  \begin{equation}
  \label{a2}
  \langle\delta(R)\delta(R')\rangle=\frac{1}{2\pi^2}\int_{0}^{\infty}k^2P(k){\widehat{W}_f}(k;R){\widehat{W}_f}(k;R')\mathrm{d}k
  .\end{equation}
  Since $S$ is a decreasing  function of $R$ and $R$ an increasing function the mass $M$ contained in the sphere of radius $R$, then, $S$ can be considered as a function of mass. \\
  The most interesting filter, because of its obvious physical meaning, is the top-hat in real space given by
  \begin{equation}
  \label{a3}
  W_f(r;R)=H\left(1-\frac{r}{R}\right)\frac{1}{\frac{4}{3}\pi R^3}
  ,\end{equation}
  where $H$ is the Heaviside step function. The Fourier transform of the filter is given by,
  \begin{equation}
  \label{a4}
  \widehat{W}(k;R)=\frac{3[\mathrm{sin}(kR)-kR\mathrm{cos}(kR)]}{k^3R^3}
  .\end{equation}
 In this paper we assume a Gaussian density field and the top-hat filter in real space. We used a flat model for the Universe with
  present day density parameters $\Omega_{m,0}=0.3$ and
   $ \Omega_{\Lambda,0}\equiv \Lambda/3H_0^2=0.7$, where
  $\Lambda$ is the cosmological constant and $H_0$ is the present day value of Hubble's
  constant. We have used the value $H_0=100~\mathrm{hKMs^{-1}Mpc^{-1}}$
  and a system of units with $m_{unit}=10^{12}M_{\odot}h^{-1}$,
  $r_{unit}=1h^{-1}\mathrm{Mpc}$ and a gravitational constant $ G=1$. {In these units},
  %  At this system of units
  $H_0/H_{unit}=1.5276.$ Regarding  the power spectrum, we  employed the $\Lambda CDM$ formula proposed by
  \citep{Smith1998}.

~\\
The stochastic process is defined as follows. We assume that
 \begin{equation}
 \label{b1}
 \delta(S)=\int_{0}^S K(S,u)\mathrm{d}W(u)
 ,\end{equation}
 where $K$ is a kernel and $\mathrm{d}W$ is the usual Wiener process. Thus, in the  plane $(S,\delta)$ we have a random walk. We assume a kernel of the simple form
 \begin{equation}
 \label{b2}
 K(S,u)=c\left[1-a\frac{u}{S}\right]
 ,\end{equation}
 for $u\leq S$ and zero otherwise. Substituting in Eq. \ref{b1} and integrating by parts we have
 \begin{equation}
 \label{b3}
 \delta(S)= c(1-a)W(s)+c\frac{a}{S}\int_{0}^{S}W(u)\mathrm{d}u
 .\end{equation}
 {Thus $\delta$ is a linear combination of a Wiener process, $W(S)$,  and an average integrated Wiener process,$\frac{1}{S}\int_{0}^{S}W(u)\mathrm{d}u$. For $a=0$ the variable  $\delta$ describes a Wiener process in which  the  value of  $\delta (S+\Delta S)$ depends only on the value of $\delta (S)$ and not on previous values, $\delta (S+\Delta S)=\delta (S)+c(1-a)\Delta W(s)$. This is because, according to the definition of Wiener process, at every step the increment $\Delta W(s)$ is chosen from a central Gaussian with variance $\Delta S$. Thus the steps of a walk on the $(S,\delta)$ plane are uncorrelated. For $a\neq 0$ the second term in the right hand side of Eq. \ref{b3}  includes information from all positions of the walk up to $S$ and results to a correlation between steps.}
 Obviously $\delta$ is a central Gaussian as a sum of central Gaussians.\\
 The autocorrelation between scales  is found multiplying $\delta(S)$ by $\delta(S')$ and finding the expected value of the product taking into account the following property of Wiener integration, see for example Jacobs (2010)
%\cite{jac},
 \begin{equation}
 \label{b4}
 \langle \int_{0}^S f(u)\mathrm{d}W(u)\int_{0}^{S'} g(u)\mathrm{d}W(u)\rangle=\int_{0}^{min\{S,S'\}}f(u)g(u)\mathrm{d}u
 .\end{equation}
  For $S'\leq S$ we have
 \begin{equation}
 \label{b5}
 \langle \delta(S)\delta(S')\rangle=c^2S'\left(1-\frac{a}{2}\right)\left[1+\frac{a(2a-3)}{3(2-a)}\frac{S'}{S}\right]
 .\end{equation}
 Then,
 \begin{equation}
 \label{b6}
 \langle \delta^2(S)\rangle=c^2S\left(\frac{a^2}{3}-a+1\right)
 .\end{equation}
 From Eqs. \ref{a0} and \ref{a1} we have the condition $c^2[a^2/3-a+1]=1$. Then, Eq. \ref{b5} can be written as
 \begin{equation}
 \label{b6b}
 \langle \delta(S)\delta(S')\rangle=S'+\lambda \frac{S'(S-S')}{S}
 ,\end{equation}
 where $\lambda=\frac{a(3-2a)}{2(a^2-3a+3)}$.\\
 It is reported in Eq. 90 of \cite{Maggiore2010a},  and is confirmed  by our calculations that for the top-hat filter the predictions of Eq. \ref{b6b} are in good agreement with those of Eq. \ref{a2} for values of $\lambda$ close to 0.5. In Fig.1 we give an example. The prediction of Eq. \ref{a2} and the prediction of  Eq. \ref{b6b} for $\lambda=0.45$ that corresponds to $a=1.185$ or $a=0.796$ are plotted. This choice for the  values of $a$, and consequently of $c$,   gives the correct correlation between scales which is very important in describing the accurate evolution of random walks. This evolution corresponds completely to the power spectrum and the smoothing filer used and first distributions which will be presented below are fully consistent with the idea of the excursion set model. Since the excursion set model and the first crossing distribution are inseparably linked, an accurate evaluation of the first crossing of a barrier  by the above random walks is essential.\\
 An interesting quantity which correlates the past the present and the future of a time evolving stochastic process $x(t)$  is defined by
 \begin{equation}
 \label{cor1}
 C(\Delta, x(t))=E\left[(x(t)-x(t-\Delta))(x(t+\Delta)-x(t))\right]
 .\end{equation}
 In our problem this quantity gives a correlation between the steps of random walks for various values of $S$. Using  Eq. \ref{b6b} we have
 \begin{gather}
  \label{cor2}
 C(\Delta, \delta(S))=E\left[(\delta(S)-\delta(S-\Delta))(\delta(S+\Delta)-\delta(S))\right]=\nonumber\\
 \frac{\Delta^2\lambda(2S-\Delta)}{S(S+\Delta).}
 \end{gather}
 Obviously {when $\lambda=0$ also $C=0$} and the steps are uncorrelated. This corresponds to a Wiener process. For positive values of $\lambda$, steps are positive correlated for $\Delta <2S$ (persisting walks) and negative correlated for $\Delta >2S$ (anti-persisting walks). We note that for the fractional Brownian motion, which is a procedure with correlated steps, the persisting case corresponds to a Hurst exponent $H>1$ and the anti-persisting  to $H<1$, (see for example \cite{Hiotelis2013}). However, roughly speaking, the procedure studied above looks like a fractional Brownian motion with varying $H$.\\
 \begin{figure}
 \includegraphics[width=9cm]{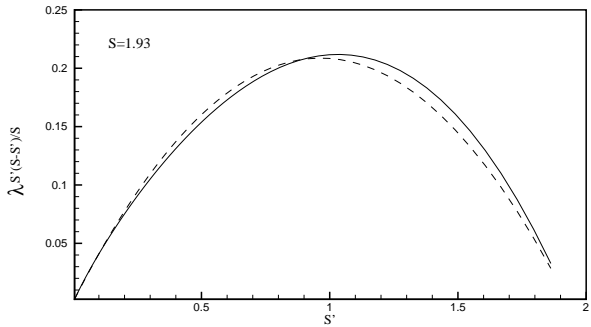}
\caption{ Predictions of Eqs. \ref{a2} and \ref{b6b}, solid line and dashed line respectively, for $\lambda =0.45$. }\label{fig1}
 \end{figure}

\section{Results and discussion}
We  discretize  Eq. \ref{b3} by dividing the {
mass interval $[M_{min},M_{max}]=[10^{-3},10^{5.5}]M_{unit}$} into $n$ intervals of equal length in logarithm spacing.  $N$ tracer particles are considered and $\Delta W_i, i=1,2..n $ values for each tracer particle are chosen from central Gaussians with respective variances $S(M_{i-1})-S(M_i)$. Then, $\delta S(i)$  is calculated according to Eq. \ref{b3}. The first crossing of the
constant barrier $\delta_c=\delta(z)$ is found for every tracer particle. We recall that $\delta(z)$ is the linear extrapolation up to present of the overdensity of a spherical region which collapses at redshift $z$ \citep{Peebles1980}. The number $n_i$ of particles which have their first upcrossing of the barrier between $S_{i-1}-S{_i}$ are grouped and the first crossing distribution is calculated by $f(S_i)=n_i/[N(S_{i-1}-S{_i})]$. Finally, the mass function is calculated by $2S_if(S_i)$.
 \begin{figure}
 \hspace{-1.5cm}
\includegraphics[width=12cm]{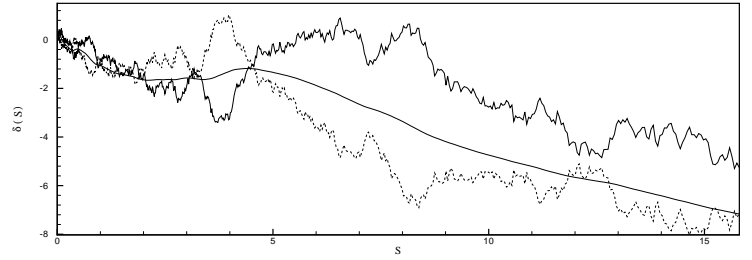}
\caption{ Role of the  kernel in amplifying the walk of a tracer particle. Smooth solid,  noisy-dashed and  noisy-solid lines correspond to $a=1$, $a=0$ and $a=3/2$ respectively. }\label{fig2}
 \end{figure}
In Fig. 2 we show the paths of the same tracer particle for  $a=1$  and for the cases $a=0$ and $a=3/2$. These paths are represented in the figure. It is clear that the path with $a=1$  is much smoother, as expected. Consequently values of $a$ control the degree of smoothness which is related to the distribution of values of $\delta (S+\mathrm{d}S)$  for given $\delta(S)$. So the values of $a$ define the correlation between various scales and, as we have shown above, a proper choice of these values results to  autocorrelation functions which approximate very satisfactory the results of Eq. \eqref{a2}
.\\
Before studying mass functions  we have checked the reliability of our  Monte Carlo approximation by testing our results with analytical solutions{.} For $a=0$ or $a=3/2$ the procedure is a Wiener process and the first crossing distribution is given by the inverse Gaussian
\begin{equation}
\label{c1}
f_{invG}(S)=\frac{\delta_c}{\sqrt{2\pi}}S^{-\frac{3}{2}}e^{-\frac{\delta^2_c}{2S}}
.\end{equation}
In Fig. 3, the prediction of Eq. \ref{c1} is {plotted together with }
%plaotted, as is 
the prediction of our Monte carlo approximation.
The horizontal axis is $\nu=\delta_c(z)/\sqrt{S}$ for $z=0$. We note that this is  a test with only numerical interest. The results presented in this figure are {derived}
%predicted 
for $n=1000$ and $N=5\times 10^5$. Our results are also compared with the interesting analytical predictions of \cite{Maggiore2010a} and \cite{Musso2012}.\\
 In \cite{Maggiore2010a} the authors use a path integral approach to estimate the first crossing distributions and their results have the form of infinite series which converge slowly. Their approximation is fully consistent  with the idea of the excursion set model. The resulting mass function is approximated by the formula,
 \begin{figure}
 \includegraphics[width=9cm]{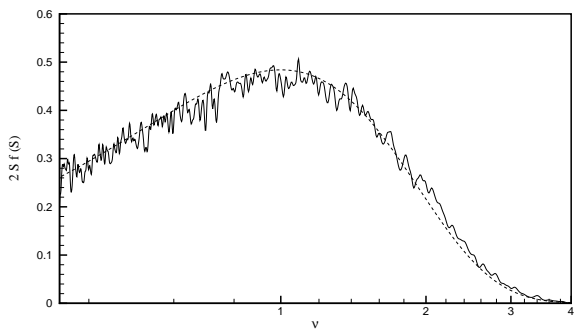}
\caption{ Comparison of the results of our Monte Carlo method with the exact solution for the  cases $a=0$ or $a=3/2$. The agreement is satisfactory.}\label{fig3}
 \end{figure}
\begin{equation}
\label{c2}
2Sf(S)=(1-\lambda)\left(\frac{2}{\pi}\right)^{1/2}\nu e^{-\frac{1}{2}{\nu}^2}+\frac{\lambda} {\sqrt{2\pi}} \nu G \left (\frac{1}{2} {\nu}^2\right)
,\end{equation}
where $G(x)=\int_{x}^{\infty}t^{-1}e^{-t}\mathrm{d}t$, (see Eq. 120 in \cite{Maggiore2010a}) .\\
On the other hand, in the approximation \cite{Musso2012} the condition of the first up-crossing  is replaced by a condition of any up-crossing, while these two conditions are obviously not equivalent. Additionally, a bivariate joint distribution between $\delta$ and $v\equiv \mathrm{d}\delta/\mathrm{d}S$ is assumed, a choice which is unjustified. The mass functions is approximated by
\begin{equation}
\label{ms1}
2Sf(S)=Sf_{invG}R(\Gamma,\nu)
,\end{equation}
where
\begin{equation}
\label{ms2}
R(\Gamma,\nu)=\frac{1+\mathrm{erf}(\Gamma \nu\sqrt{2})}{2}+\frac{e^{-\frac{1}{2}\Gamma^2\nu^2}}{\sqrt{2\pi}\Gamma\nu}
,\end{equation}
where $\Gamma$ depends on the power spectrum and the kernel used to smooth the density field.

 \begin{figure*}
 \includegraphics[width=18cm]{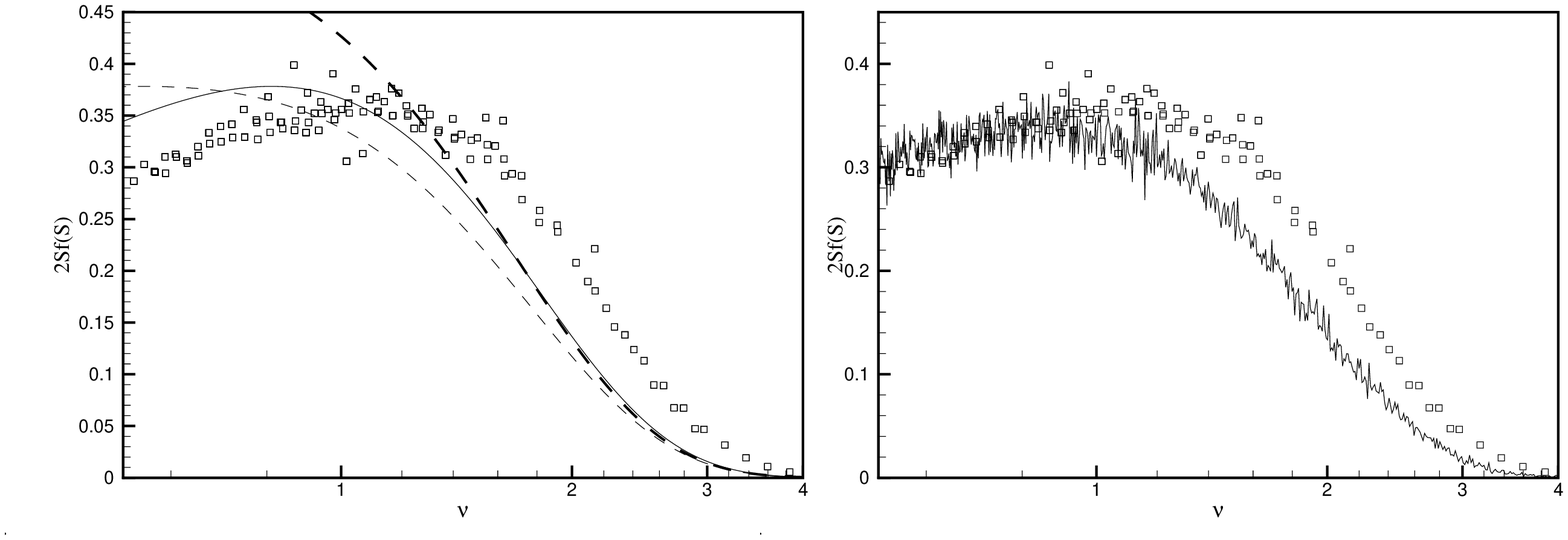}
\caption{ Comparison of the results of N-body simulations, squares, with those of  analytical formulae of Eq. \ref{c2} and Eq.  \ref{ms1}, and with our predictions. Left snapshot: {squares}
 are the results of N-body simulations. The predictions of Eq. \ref{c2} are represented by the smooth solid line  while the predictions of  Eq. \ref{ms1} for $\Gamma=1/3$ and $\Gamma=1/2$ are represented by the dashed lines (large dashes and small dashes respectively). Right snapshot: {squares} are the results of N-body simulations. The results of our Monte Carlo simulations are given by the solid line. }\label{fig4}
 \end{figure*}
 \begin{figure}
 \hspace{-1.5cm}
 \includegraphics[width=12cm]{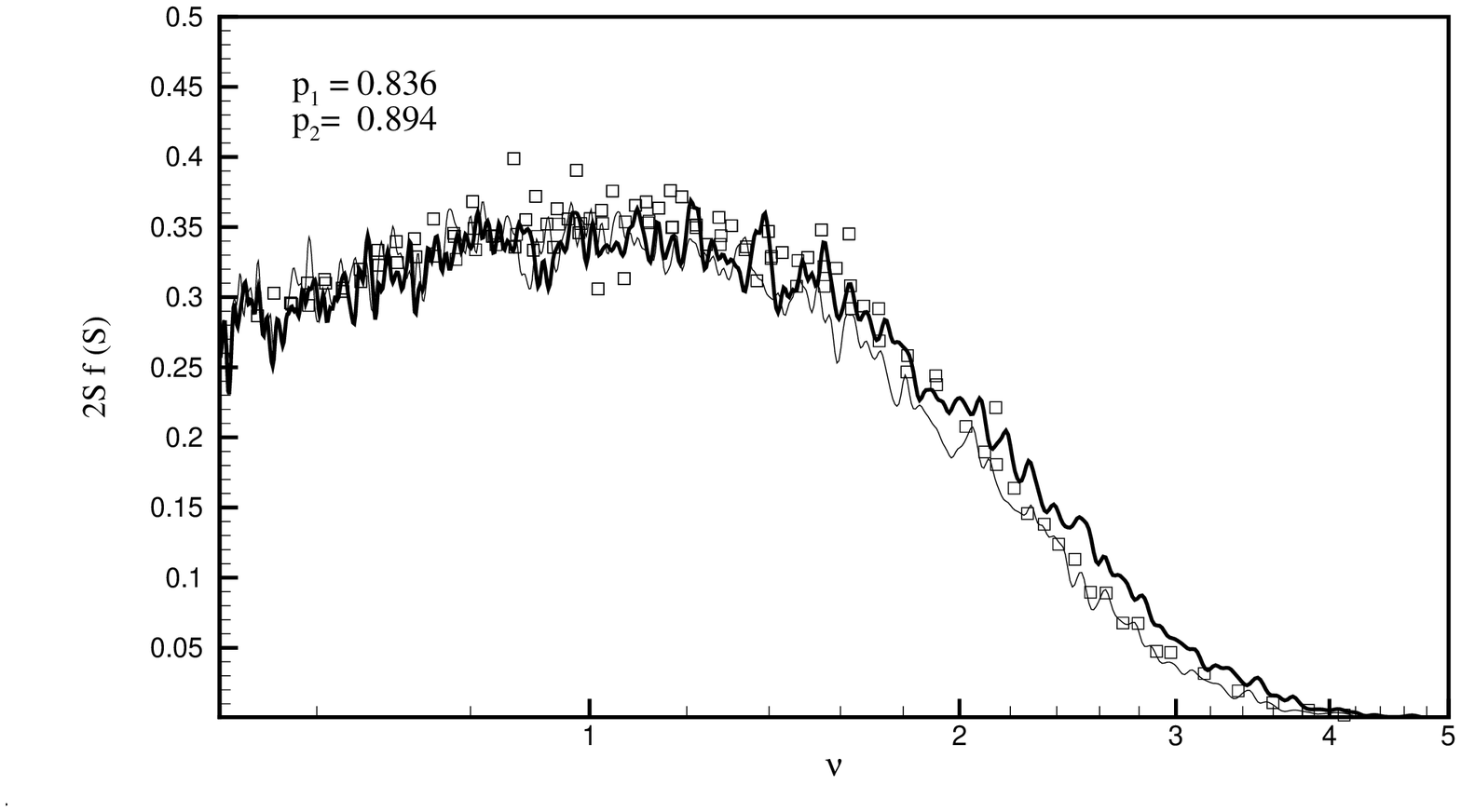}
\caption{ Comparison of the results N-body simulations, squares, with our results. The thin  solid line solid  corresponds to $p=0.894$ and the thick one to $p=0.836$. }\label{fig5}
 \end{figure}
In Fig. 4, we compare the predictions of our results, {derived}
for $a=0.796$,  with those of  analytical formulae of Eq. \ref{c2} and Eq. \ref{ms1}  with the predictions of N-body simulations at $z=0$. In both snapshots, squares are the predictions of N-body simulations of \cite{Tinker2008}.

The prediction of Eq. \ref{c2} and the predictions of Eq. \ref{ms1} for $\Gamma=1/3$  and for $\Gamma=1/2$  are plotted in the figure.  In the right snapshot the prediction of Monte Carlo approximation. Analytical formulae of Eq.\ref{c2} and Eq. \ref{ms1} result to smaller numbers for heavy haloes and larger numbers for smaller haloes compared to the results of N-body simulations, while our approximation gives the correct behavior of the mass function for small haloes, $\nu\leq 1$. This is an interesting result. It shows that the excursion set model works very satisfactory for small haloes with $\nu \leq 1$. This agreement has not been reported  elsewhere. On the other hand, in agreement with the analytical formulae studied above, our approximation fails to produce the correct number of heavier haloes but for $\nu \geq 1.1$ our results coincide with those of 
Eq. \ref{c2} and those of Eq. \ref{ms2} (for $\Gamma$ =1/3). Definitely, the problem of the approximation of the correct first crossing distribution by a simple analytical formula has not been solved yet but we believe that the simplicity of the approximation formula is a secondary issue. The important issue is that of the accurate  evaluation of the first crossing distributions at various scales. Our results and those of the analytical formulae of Eq. \ref{c2} and Eq. \ref{ms1} indicate that the predictions of the excursion set model are, in any case,  for heavy haloes far from the results of N-body simulations at least for the case of the top-hat filter and the constant barrier.\\
    It seems possible that at large scales, additional parameters  may be taken into account, as for example the ellipticity or the angular momentum of the structures. This could lead to {think that}
%the thought  that 
    the use of moving barriers as those proposed in the literature,
\citep{Bond1996a,DelPopolo1998,Sheth2001,Sheth2002} is necessary. Moving barrier models try to solve the problem of the overestimation of the number of small haloes and the underestimation of the number of small haloes using  a critical threshold of collapse which varies with mass ($S$). The choice of a moving barrier is based on physical arguments since larger haloes appear with the larger ellipticity and larger angular momentum. A smaller critical threshold of collapse for large  haloes  rearranges  first crossing distributions  at various scales. Studying various cases of moving barriers we found that an agreement with the predictions of N-body simulations can be achieved  using a constant, lower threshold for collapse. We used  a barrier of the form $\delta_*(z)=p\delta_c(z)$ where $p$ is a constant.  In Fig.5 we present a comparison  with the results of N-body simulations for two different values of $p$ at $z=0$. An agreement is shown. It is interesting to note the sensitivity of the distribution of heavy haloes to the values of $p$. We note that the use of a lower threshold results to an increase of the fraction $n_{cross}/N$ where $N$ is the total number of walks studied and $n_{cross}$ is the number of walks which have passed the threshold in the range $S_{min}-S_{max}$, but this increase is larger for small values of $S$ (larger haloes). \\
\begin{figure}
 \includegraphics[width=10cm]{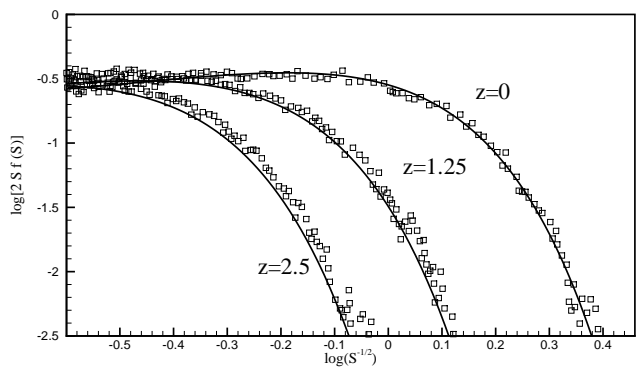}
\caption{ Comparison of our results, open squares  with those of N-body simulations. The results of N-body simulations are represented by the fitting formula of \protect\cite{Tinker2008} and are shown by the thick solid lines. Our results have been {derived} for $a=0.796$ and $p=0.866$.}\label{fig6}
 \end{figure}
 We used  $p=0.866$ as the best value. 
%{\bf SENTENCE TO BE CUT Thus in what follows we use the term out results in order to denote those which have %been derived for $a=0.796$ and $p=0.866$}. 
We calculated mass functions for  redshifts  $z=0, z=1.25$ and $z=2.5$ and we presents them {\bf in} Fig. 6. Our results are plotted {in } the figure as are those derived from the fitting formula of \cite{Tinker2008} given in their Eq. 3  which is
 \begin{figure}
 \includegraphics[width=10cm]{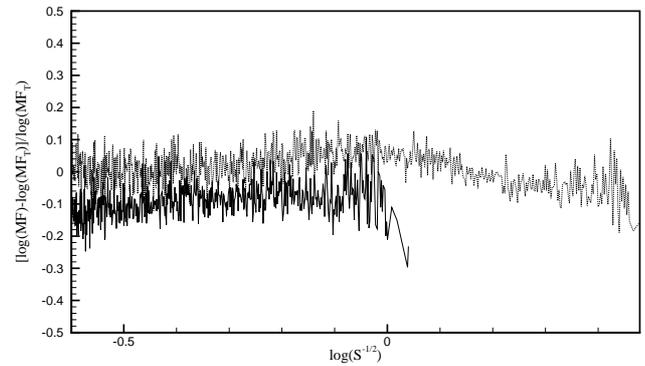}
 \hspace{-2.5cm}
\caption{ Fractional error $[\log(MF)-\log(MF_T)]/\log(MF_T)$ for two redshifts $z=0$ {(dotted-line)}
 and $z=2.5$ {(solid-line)}. $MF_T$ is {given} by the fitting formula of \protect\cite{Tinker2008} while $MF$ represents our results.}\label{fig7}
 \end{figure}
 \begin{figure}
 \hspace{-1.5cm}
 \includegraphics[width=10cm]{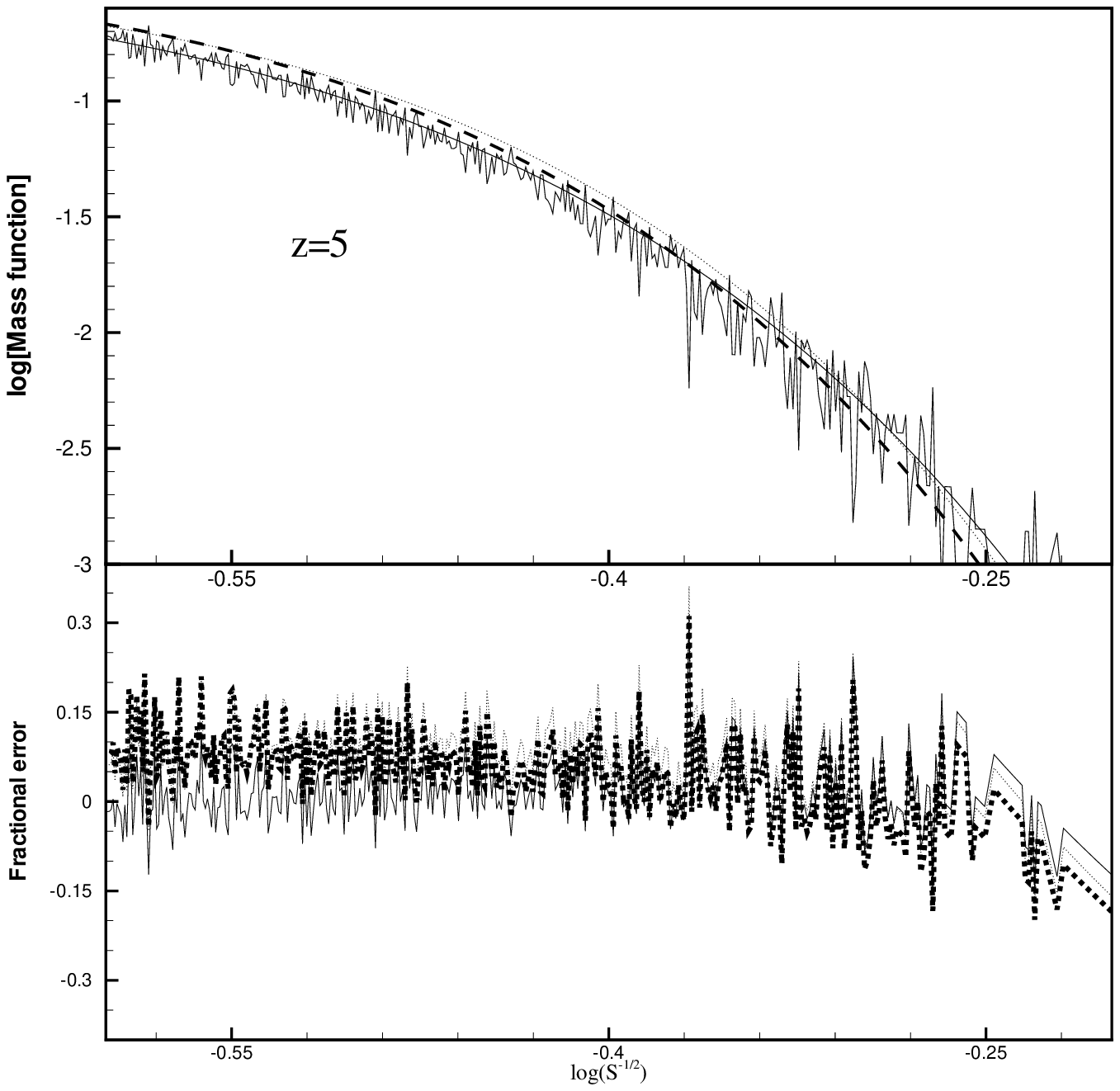}
 \hspace{2.5cm}
\caption{ Upper snapshot: Mass functions at redshift $z=5$. The noisy solid line shows our results ($a=0.796, p=0.866$). The smooth solid line shows the model of \cite{Sheth2001}. Dashed line represents the results of  \cite{Warren2006} while dotted line shows the results of  \cite{Watson2013}.\newline
Lower snapshot: Fractional errors between our results and the model of \cite{Sheth2001} , solid line, between our results and the model of \cite{Warren2006}, dashed line, and between our results and the model of  \cite{Watson2013}, dotted line.}\label{fig8}
 \end{figure}
 \begin{equation}
 \label{tinkform}
 MF_T(\sigma,z)=A\left[\left(\frac{b}{\sigma}\right)^{a}+1\right]e^{-\frac{c}{\sigma^2}}
 ,\end{equation}
 where  $\sigma=\sqrt{S}$ and the z-dependence is given by
 \begin{gather}
 A=0.186(1+z)^{-0.14},~~
 a=1.47(1+z)^{-0.06},\nonumber\\
 b=2.57(1+z)^{-\alpha},~~
 c=1.19
 \end{gather}
 \begin{equation}
 \label{tinkforma}
  \alpha=\exp\left[-\left(\frac{0.75}{\log(\Delta_{vir}/75)}\right)^{1.2}\right]
 ,\end{equation}
 ( See Eqs 5,6,7, and 8 in  \cite{Tinker2008}). We used $\Delta_{vir}=200$. \\
 In Fig.7 we show the fractional error, defined by $\mathrm{fr_{error}}=\frac{\log(MF)-\log(MF_T)}{\log(MF_T)}$ where $MF$ is the mass function derived by our model, for $z=0$ and $z=2.5$.\\
  {We note that according to \cite{Tinker2008} their model described by  Eq. \eqref{tinkform} is valid for $0\leq z\leq 2.5$. However in order to find the cause of increasing difference between our results and those of \cite{Tinker2008}, shown in Fig.7, we present comparisons with some other analytical mass functions available in the literature, {such that of \\ 
%  The formula of  \\
  A. \cite{Watson2013} which is valid for $0\leq z \leq 30$. This is given by
 \begin{equation}
 MF_{Wats}=MF_T
 ,\end{equation}
 where
 \begin{equation}
 A=0.282,~~ a=2.163,~~ b=1.406,~~ c=1.21
 .\end{equation}
 \\
 B. The formula of \cite{Warren2006}, which is,
 \begin{equation}
 MF_{War}=0.7234(\sigma^{-1.625}+0.2538)e^{-\frac{1.1982}{\sigma^2}}
 ,\end{equation}
 and 
 \\
 C. that of \cite{Sheth2001}
 }
 \begin{equation}
 MF_{ST}=A\sqrt{\frac{2a_s}{\pi}}\left[1+\left(\frac{\sigma^2}{a_s\delta_c^2}\right)^{p_{s}}\right]
 \frac{\delta_c}{\sigma}e^{-\frac{a_s\delta_c^2}{2\sigma^2}}
 ,\end{equation}
 where $A=0.3222,~a_s=0.707$ and $p_s=0.3$.\\
 The comparisons for  redshift $z=5$ are shown in Fig. 8. We show that the agreement remains satisfactory. \\
  We also note that for large redshifts resolution problems appear. This is because  $\delta_c(z)$ is an increasing function of $z$ and thus the percentage  of walks which pass the barrier becomes smaller for large $z$. Large structures are more rare and the mass function appears noisy (see at the right side of Fig.8). However, it becomes difficult to check if a disagreement is due to the physical process or to the poor resolution.
 This difficulty rises the challenge of finding an analytical solution for the first crossing distribution of the process of Eq. \ref{b3}. }
 
 In Fig. 9 we present a comparison of our results with the formula of  \cite{Watson2013} and for $z=10$. These results are {derived} for $n=400$ and $N=5\times 10^6$. Resolution problems are obvious since only $4365$ from N tracer particles cross the barrier, but the agreement remains satisfactory.\\
 \begin{figure}
 \hspace{-1.5cm}
 \includegraphics[width=12cm]{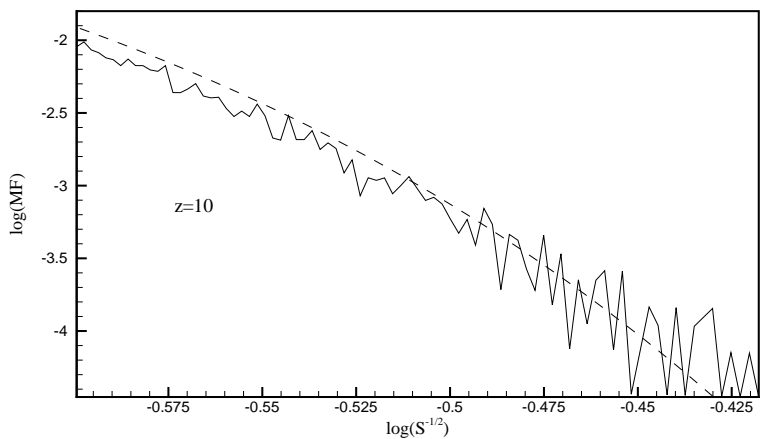}
\caption{ Mass functions at redshift $z=10$. Noisy solid line shows our results ($a=0.796, p=0.866$). The dashed line shows the model of \cite{Watson2013}.}\label{fig9}
 \end{figure}
  It is well known that the process of structure formation is a very {complex} one. It has been studied extensively in the literature and more than thirteen formulae for the mass function has been proposed for various cosmological models, various halo finding algorithms and various mass scales, see for example \cite{Watson2013} and references therein. Consequently, the probability  of constructing an analytical approach that predicts  the results of N-body simulations is extremely small. However, our results show that the stochastic process {studied here, is not a  N-body simulation, which is however able to shed} more light to the physical process during the formation of structures. Since it is a process which describes accurately the correlation between scales for the realistic top-hat filter and is able to produce results close to these of N-body simulations for a constant barrier, deserves a more profound study. Any alternative approximation of  first crossing distributions, resulting from the above described stochastic process, should be very interesting.

\section{Acknowledgements}
\ We acknowledge  Dr. Andromachi Koufogiorgou for her kind help and J.Tinker for making available the results of their N-body simulations.
ADP was supported by the Chinese Academy of Sciences and by the President’s international fellowship initiative, grant no. 2017 VMA0044.

\end{document}